\documentclass[prc,twoside,superscriptaddress,showpacs,twocolumn]{revtex4}

\usepackage{graphicx, latexsym, amssymb, amsmath, color, multirow, mathrsfs, CJK, ifpdf}
\usepackage[section]{placeins}
\usepackage{amsmath}
\usepackage{amsfonts}
\usepackage{amssymb}
\usepackage{bm}
\usepackage{graphicx}
\usepackage{txfonts}
\usepackage{ulem}

\newcommand{\delete}{\bgroup\markoverwith{\textcolor{red}{\rule[0.5ex]{2pt}{1pt}}}\ULon}

\makeatletter

\newcommand{\Rmnum}[1]{\expandafter\@slowromancap\romannumeral#1@}
\makeatother

\begin{document} 
\preprint{APS/123-QED}

\title{Tensor effects on gap evolution of $N=40$ from non-relativistic and relativistic mean-field theory}
\author{Long Jun Wang }
\affiliation{School of Nuclear Science and Technology, Lanzhou University, Lanzhou 730000, China}
\author{Jian Min Dong }
\affiliation{Institute of Modern Physics, Chinese Academy of Science, Lanzhou 730000, China}
\author{Wen Hui Long }
\email{longwh@lzu.edu.cn}
\affiliation{School of Nuclear Science and Technology, Lanzhou University, Lanzhou 730000, China}

\date{\today}
\begin{abstract}
Tensor effects on the $N=40$ gap evolution of $N=40$ isotones are studied by employing the Skyrme-Hartree-Fock-Bogoliubov (SHFB) and relativistic Hartree-Fock-Bogoliubov (RHFB) theories. The results with and without the inclusion of the tensor components are compared with each other. It is found that when the tensor force is included, both two different approaches present the same trend and qualitatively agree with the experimental one indicated by the $\beta$-decay studies, which implies an important role played by the tensor force in the gap evolution of $N=40$. Furthermore, it is shown that the gap evolution is dominated by the corresponding tensor contributions from $\pi$ and $\rho$-tensor coupling in the relativistic framework.
\end{abstract}


\maketitle 

With worldwide development of Radioactive Ion Beam (RIB) facilities, the nuclei far from the valley of $\beta$-stability, namely the exotic nuclei, have become more accessible in recent years \cite{Exotic2000PPNP}. The investigations on the shell structures of such nuclei have formed a new frontier in modern nuclear physics due to some novel phenomena discovered timely. The typical examples are the emergence of some new magic numbers \cite{Ozawa2000PRL, Liddick2004PRC} and the quenching of some conventional shell closures \cite{Ozawa2000PRL, Kowalska2006HI, Caurier2004NPA} when approaching the isospin limits of the realistic nuclei. As suggested by Otsuka and the collaborators, the tensor force may be one of the crucial physical mechanisms in modeling the shell evolution and the occurrence of magic numbers in the exotic regions \cite{Otsuka2005PRL, Otsuka2006PRL, Otsuka2010PRL}. However, in both non-relativistic and relativistic self-consistent mean-field calculations, the relevant tensor components were dropped up to very recently \cite{Brown2006PRCR, Lesinski2007PRC, Bender2009PRC, Long2006PLB1, Long2007PRC, Lalazissis2009PRCR}.

Since the magic character of $^{68}$Ni was proposed in the early eighties by Bernas \textsl{et al.} \cite{Bernas1982PLB}, the subshell closure of $N=40$ has been devoted more and more efforts from both experimental and theoretical sides due to the significant roles played by the nuclei of $N = 40\sim60$ in the $r$- and $rp$-process paths \cite{Kanungo2007PLB}. In the shell-model picture, the gap of $N=40$ is determined by the negative parity $fp$ ($2p_{1/2}$ and $1f_{5/2}$) and positive parity $1g_{9/2}$ orbits. Thus, the low-lying structure of $^{68}$Ni might be concerned with the change of parity \cite{Sorlin2002PRL}. On the other hand, due to the opposite tensor effects felt by the $1f_{5/2}$ (or $2p_{1/2}$) and $1g_{9/2}$ states, the tensor force is also expected to play an important role in figuring the $N=40$ gap evolution since most of the $N=40$ isotones are spin-unsaturated for proton \cite{Otsuka2005PRL, SHF2007PRC}.  Recently several experiments of $\beta$-decay indicate a rapidly weakening of $N=40$ gap when removing protons from $^{68}$Ni \cite{Mueller1999PRL, Sorlin2001NPA}, whereas such feature cannot be properly described by the mean-field calculations when the tensor components of the nuclear interaction are excluded \cite{Gaudefroy2009PRC, Nakada2010PRCR}.

In this Brief Report, the gap evolution of $N=40$ will be explored within the non-relativistic and relativistic mean field theories, specifically the non-relativistic Skyrme-Hartree-Fock-Bogoliubov (SHFB) \cite{Skyrme1956PM}, the relativistic Hartree-Bogoliubov (RHB) \cite{Vretenar2005PhysRep, Meng2006PPNP}, and relativistic Hartree-Fock-Bogoliubov (RHFB) \cite{Long2010PRC} theories. Particular attention is paid to the effects of tensor couplings embedded in the neutron-proton interactions. Aiming at the isospin evolution behavior in the neutron drip-line region, we select six neutron-rich isotones of $N=40$ from $Z=20$ to $30$, which are spherical or nearly spherical as predicted by both macroscopic-microscopic model \cite{Moller1995ADNDT} and self-consistent mean-field theory \cite{Gaudefroy2009PRC}. As done by Nakada with the semi-realistic mean-field calculations \cite{Nakada2010PRCR}, the spherical calculations are performed in this study to present our qualitative understanding on tensor effects in the gap evolution of $N=40$, although some of the selected isotones might be deformed according to the shell model calculations \cite{Caurier2002EPJA}. In both non-relativistic and relativistic cases, the pairing correlations are treated within the Bogoliubov scheme such that the continuum effects can be naturally involved, which is meaningful to provide self-consistent descriptions for the selected isotones.

Within the SHFB theory, the tensor interaction is introduced as \cite{Skyrme1956PM}
\begin{align}
v_{T} =& \frac{T}{2} \bigg[\bigg( (\bm{\sigma}_1\cdot\bm{k}^\prime)(\bm{\sigma}_2\cdot\bm{k}^\prime) - \frac{1}{3}(\bm{\sigma}_1\cdot\bm{\sigma}_2)\bm{k}^{\prime2}\bigg) \delta(\bm{r}_1-\bm{r}_2) \bigg]  \nonumber \\
& +\frac{T}{2}\delta(\bm{r}_1-\bm{r}_2)\bigg[\bigg( (\bm{\sigma}_1\cdot\bm{k})(\bm{\sigma}_2\cdot\bm{k}) - \frac{1}{3}(\bm{\sigma}_1\cdot\bm{\sigma}_2)\bm{k}^{2}\bigg)\bigg] \nonumber \\
& +U\bigg[(\bm{\sigma}_1\cdot\bm{k}^\prime)\delta(\bm{r}_1-\bm{r}_2)(\bm{\sigma}_2\cdot\bm{k})-\frac{1}{3}(\bm{\sigma}_1\cdot\bm{\sigma}_2) \nonumber \\
& \times (\bm{k}^\prime\cdot\delta(\bm{r}_1-\bm{r}_2)\bm{k}\bigg], \label{tensorNR}
\end{align}
where the right operator $\bm{k}=(\overrightarrow{\nabla}_1-\overrightarrow{\nabla}_2)/(2i)$ and the left one $\bm{k}^\prime= -(\overleftarrow{\nabla}_1- \overleftarrow{\nabla}_2)/(2i)$. The coupling constants $T$ and $U$ denote the strength of the triplet-even and triplet-odd tensor components, respectively. 

Notice that the tensor force of Eq. (\ref{tensorNR}) is in a non-relativistic form, inconsistent with the Lorentz covariance. Within the relativistic framework, the corresponding tensor correlations in $T=0$ channels can be naturally taken into account by the one pion exchange and $\rho$-tensor couplings \cite{Long2007PRC, Long2008EPL}. Within RHFB, the $\pi$ and $\rho$-tensor couplings are adopted as \cite{Long2007PRC}
\begin{align}
H_\phi=& \frac{1}{2}\int d\bm{r}d\bm{r}^\prime \bar{\psi}(\bm{r})\bar{\psi}(\bm{r}^\prime) \Gamma_\phi(\bm{r},\bm{r}^\prime) D_\phi(\bm{r},\bm{r}^\prime) \psi(\bm{r}^\prime)\psi(\bm{r}),
\end{align}
where $\phi$ denotes the pseudo-vector $\pi$ and tensor $\rho$ couplings, $D_\phi(\bm{r},\bm{r}^\prime)$ is the Yukawa propagator, and the interacting matrices $\Gamma_\phi(\bm{r},\bm{r}^\prime)$ read as,
\begin{subequations}
\begin{align}
\Gamma_\pi(\bm{r},\bm{r}^\prime) \equiv& -\frac{1}{m^2_\pi}(f_\pi \vec{\tau}\gamma_5\gamma_\mu\partial^\mu)_{\bm{r}} \cdot (f_\pi\vec{\tau}\gamma_5\gamma_\nu\partial^\nu)_{\bm{r}^\prime},  \\
\Gamma^T_\rho(\bm{r},\bm{r}^\prime) \equiv& \frac{1}{4M^2}(f_\rho\sigma_{\nu k}\vec{\tau}\partial^k)_{\bm{r}} \cdot (f_\rho\sigma^{\nu l}\vec{\tau}\partial_l)_{\bm{r}^\prime}.
\end{align}
\end{subequations}
In the above equations $M$ $(m_\pi)$ denotes the mass of the nucleon ($\pi$ meson) and $f_\pi$ $(f_\rho)$ is the coupling constant of $\pi$ $(\rho)$ meson which is exponentially density dependent \cite{Long2007PRC}. Here what should be clarified is that the tensors expressed above are associated with the Lorentz rotation, different from the quantities defined by the space rotation, e.g., the ones in Eq. (\ref{tensorNR}). In fact it has been illustrated that the $\pi$ and $\rho$-tensor couplings, particularly the corresponding tensor parts with non-relativistic reduction, play a crucial role in the self-consistent description of shell structure and the evolution \cite{Long2007PRC, Lalazissis2009PRCR, Long2008EPL}.

To evaluate the tensor effects in modeling the gap evolution of $N=40$, different effective interactions are utilized in this work. For the SHFB theory, we adopt the effective interaction SLy5 with tensor components added perturbatively (marked by SLy5+T) \cite{SLy5, Colo2007PLB}, which can well reproduce the isospin dependence of the experimental energy differences between the single-proton orbits beyond the $Z=50$ core for Sn isotopes and the single-neutron ones beyond the $Z=82$ core for $Z=82$ isotones, as well as the evolution of the spin-orbit splittings of the Ca isotopes and $N=28$ isotones \cite{Colo2007PLB, Zou2008PRC}. For comparison the parameter set T31 including the tensor force (T31+T) \cite{SHF2007PRC} is also applied in the present study, which can reproduce the isospin evolution of the experimental gap between the single-proton orbits $1h_{11/2}$ and $1g_{7/2}$ along the Sn isotopic chain as well \cite{Dong2011PRC}. In the RHFB calculations we choose the effective interactions PKO2 \cite{Long2008EPL}, PKO3 \cite{Long2008EPL} and PKA1 \cite{Long2007PRC}, comparing to those by RHB with DD-ME2 \cite{Lalazissis2005PRC} in which the Fock terms are dropped. It should be also mentioned that the effective Lagrangian PKO2 does not contain $\pi$ or tensor $\rho$ couplings, whereas the former is included in PKO3, and both are taken into account by PKA1.

\begin{figure}
\ifpdf
\includegraphics[width=0.45\textwidth]{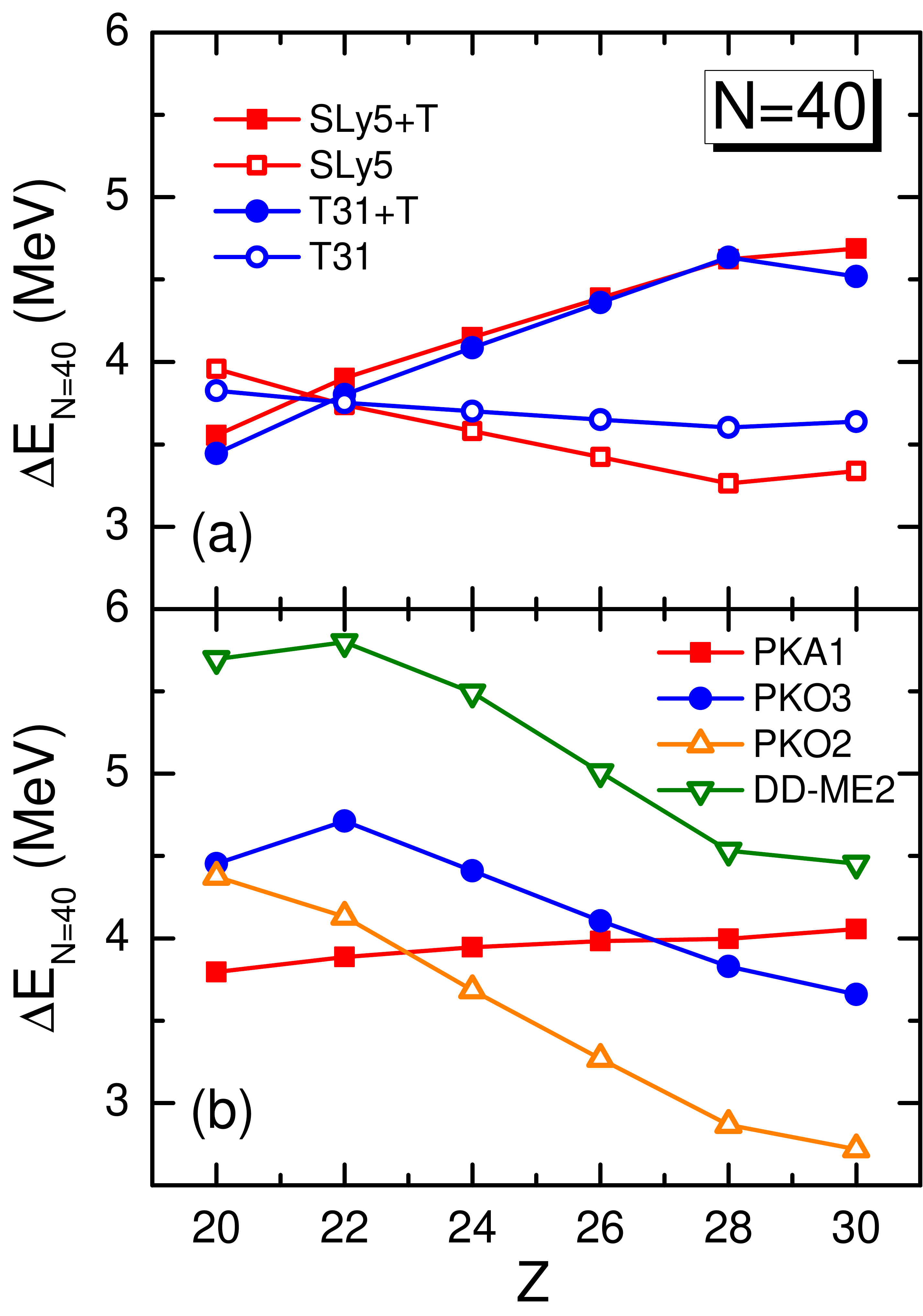}
\else
\includegraphics[width=0.45\textwidth]{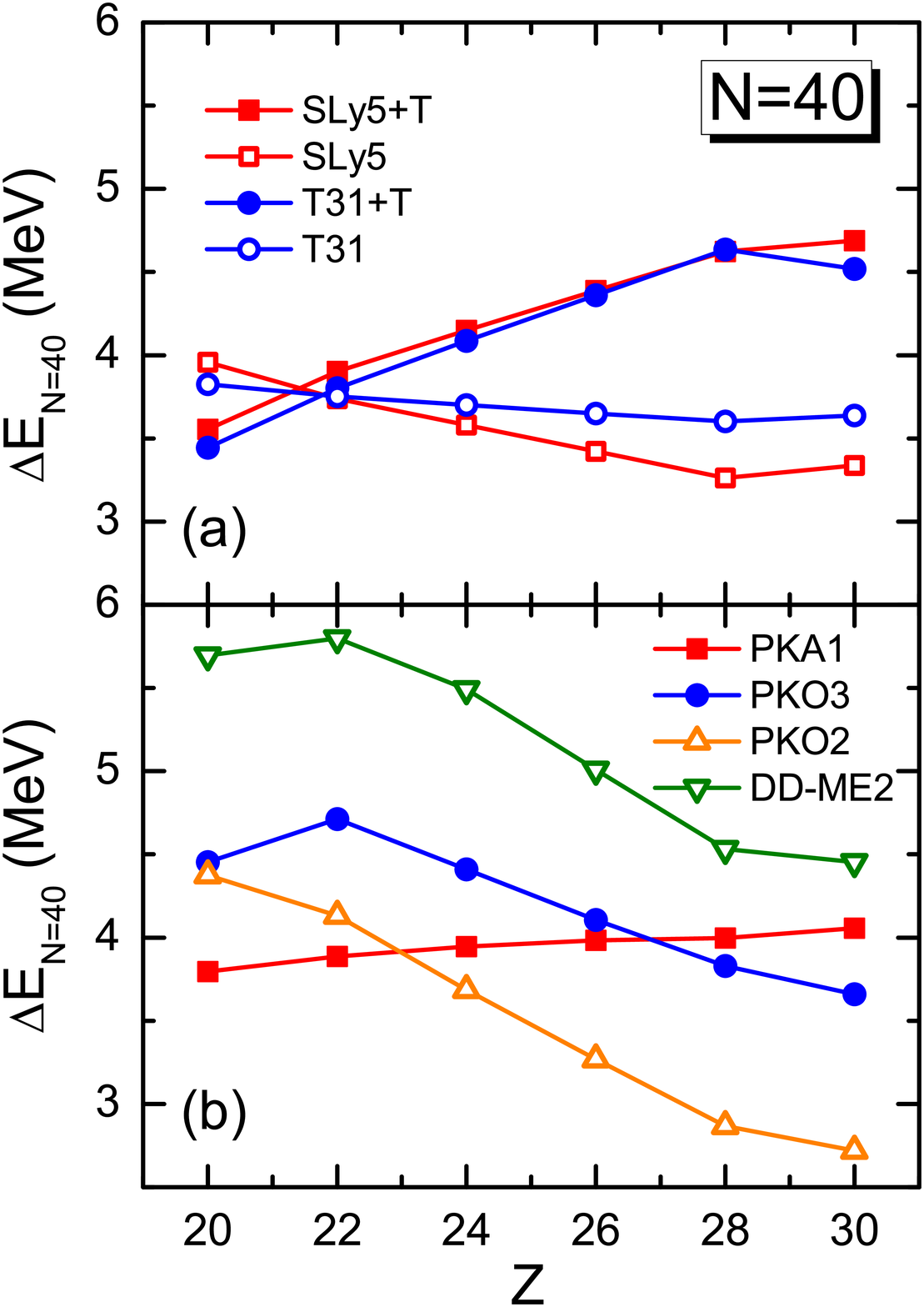}
\fi
\caption{\label{fig:fig1} (Color online) The neutron gap at $N=40$ of $N=40$ isotones as functions of the proton number $Z$. The results are calculated by the non-relativistic (a) and relativistic (b) mean-field theory with and without the tensor component. See the text for details.}
\end{figure}

Fig. \ref{fig:fig1} displays the gap evolutions of $N=40$ for the selected isotones, extracted from the calculations of the non-relativistic [plot (a)] and relativistic [plot (b)] mean field theories. Experimentally, the $\beta$-decay studies indicate that the gap of $N=40$ tends to be smaller when removing protons from $^{68}$Ni \cite{Mueller1999PRL, Sorlin2001NPA}. From Fig. \ref{fig:fig1}(a) one can see clearly that such trend cannot be reproduced by the Skyrme forces SLy5 and T31 in which the tensor components are not included. However, when the tensor force participates in the neutron-proton interaction, i.e., SLy5+T and T31+T, the calculations present identical systematics with the experiments of $\beta$-decay. Such distinct improvement can be well understood from the nature of tensor force \cite{Otsuka2005PRL}. In the SHFB calculations, the gap of $N=40$ is determined by the neutron orbits $\nu f_{5/2}$ (or $\nu p_{1/2}$) and $\nu g_{9/2}$, i.e., the ones with $j_{<}=l-1/2$ and $j_{>}=l+1/2$, respectively. Along the isotonic chain from $^{60}$Ca to $^{68}$Ni, the valence protons will gradually occupy the proton orbit $\pi 1f_{7/2}$ with $j_>=l+1/2$, which presents repulsive tensor couplings with the orbit $\nu g_{9/2}$ ($j_{>}=l+1/2$) and attractive ones with $\nu f_{5/2}$ (or $\nu p_{1/2}$). As a result the gaps are remarkably weakened by the tensor force when moving away from $^{68}$Ni, in accordance with the $\beta$-decay experiments. As compared to the SHFB calculations excluding tensor components it is then well demonstrated for the crucial role played by the tensor force in modeling the shell evolution \cite{Colo2007PLB, Dong2011PRC}.

\begin{figure}
\ifpdf
\includegraphics[width=0.45\textwidth]{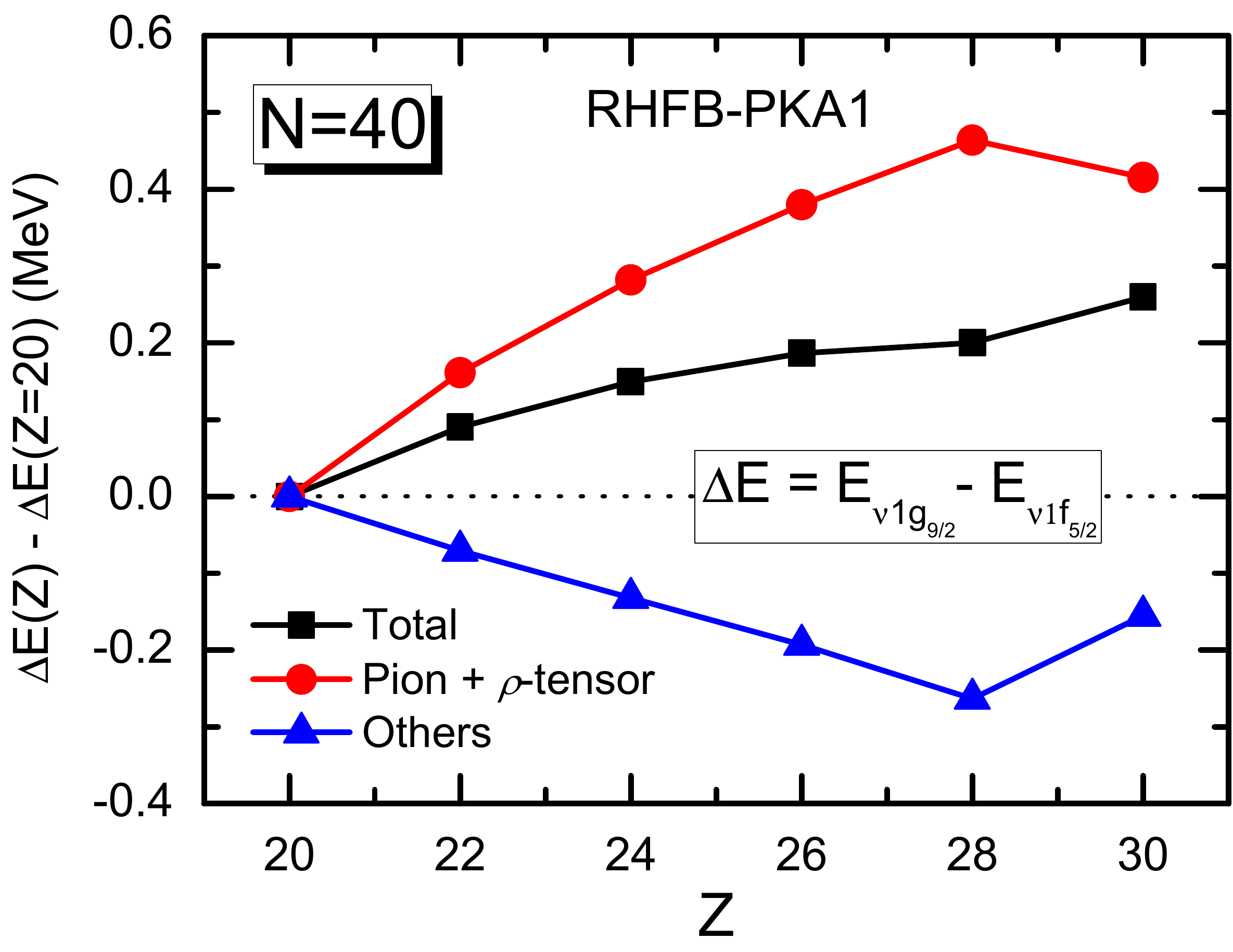}
\else
\includegraphics[width=0.45\textwidth]{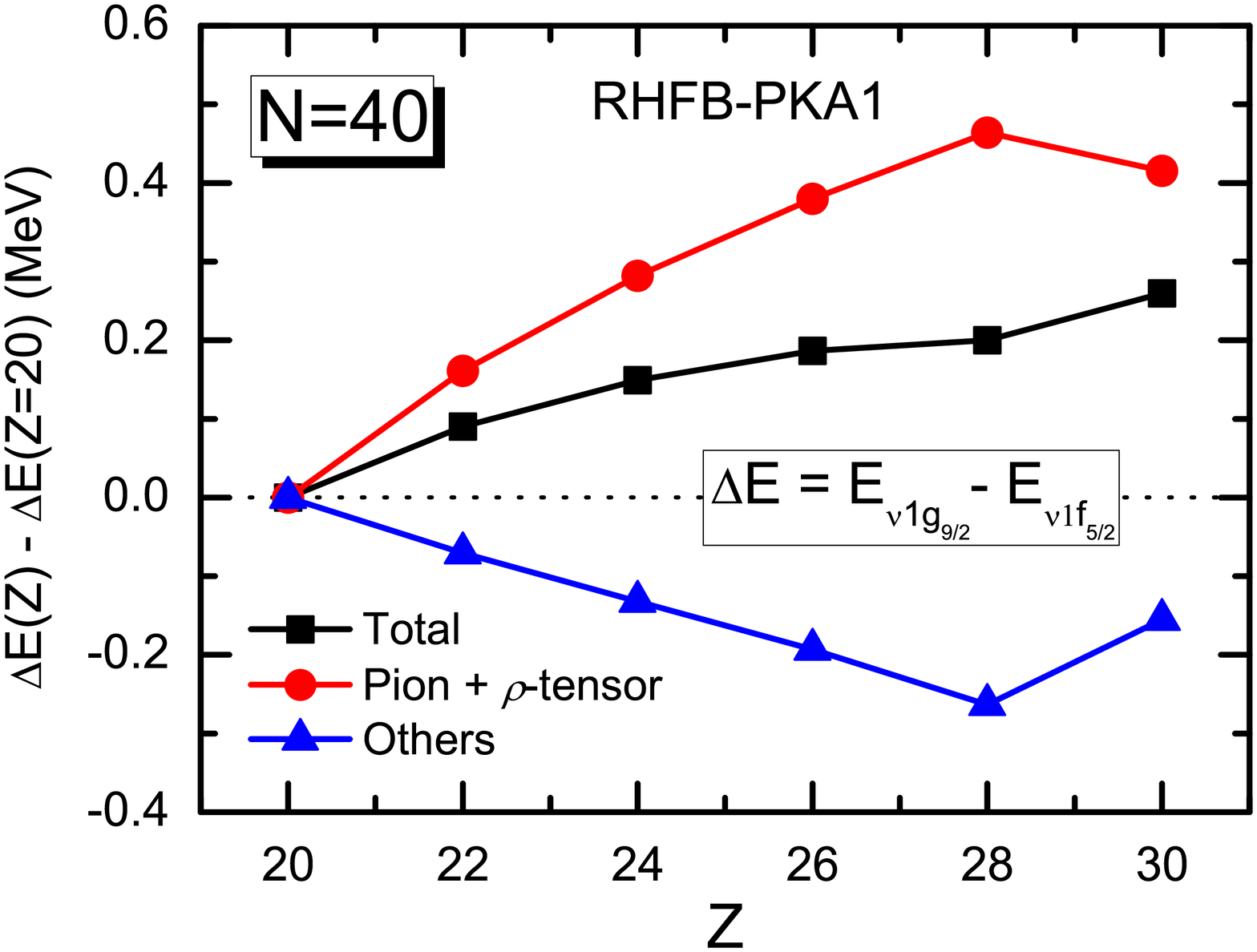}
\fi
\caption{ (Color online) Detailed contributions of the neutron gap at $N=40$ along $N=40$ isotones from the isovector $\pi$ and tensor $\rho$ couplings in comparison with those from the other channels associated with the kinetic energy, $\sigma$-scalar, $\omega$-vector, $\rho$-vector couplings, and the rearrangement term. The results are extracted from the RHFB calculations with PKA1.}\label{fig:fig2}
\end{figure}

\begin{figure}[htbp]
\ifpdf
\includegraphics[width = 0.45\textwidth]{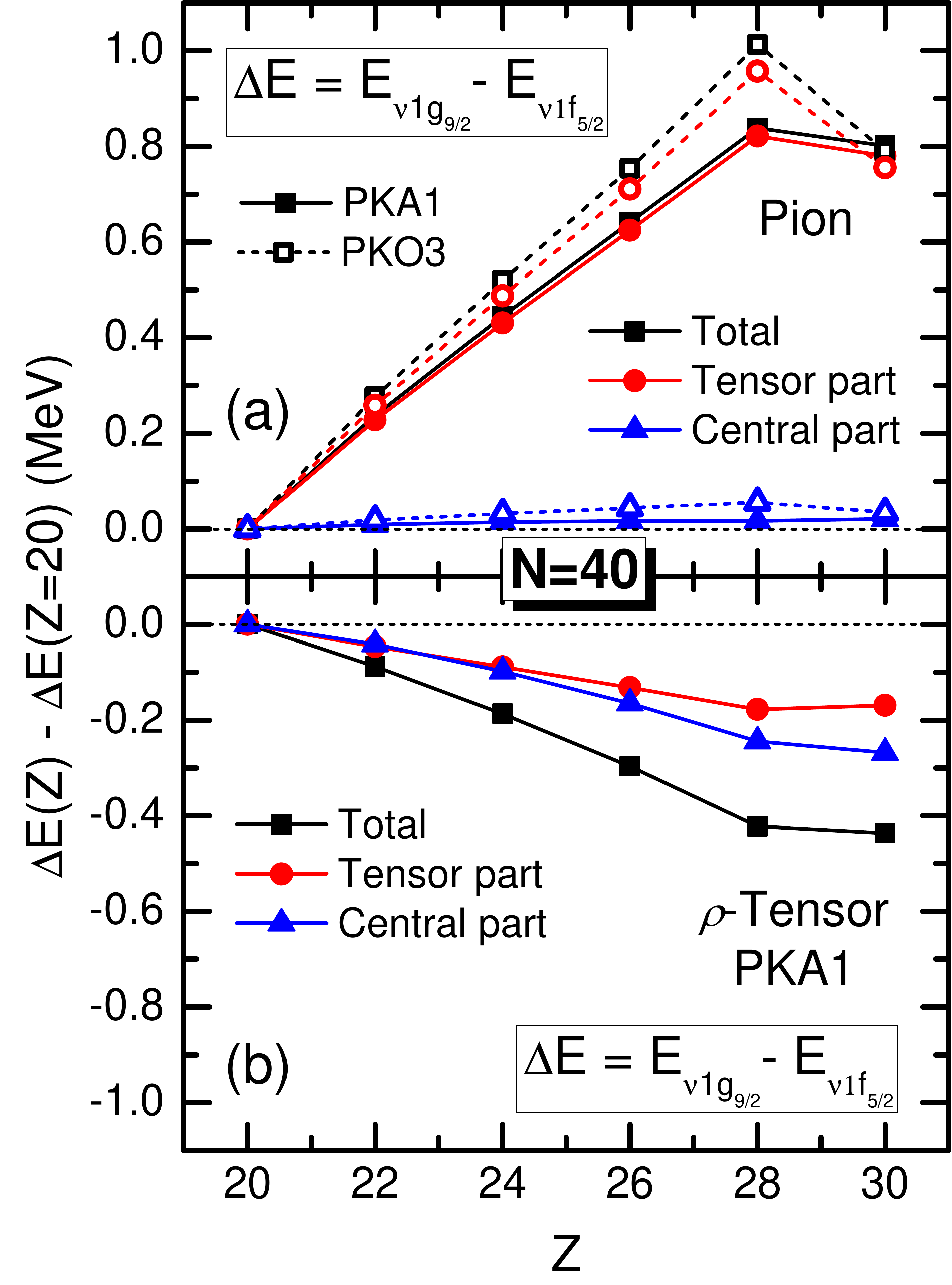}
\else
\includegraphics[width = 0.45\textwidth]{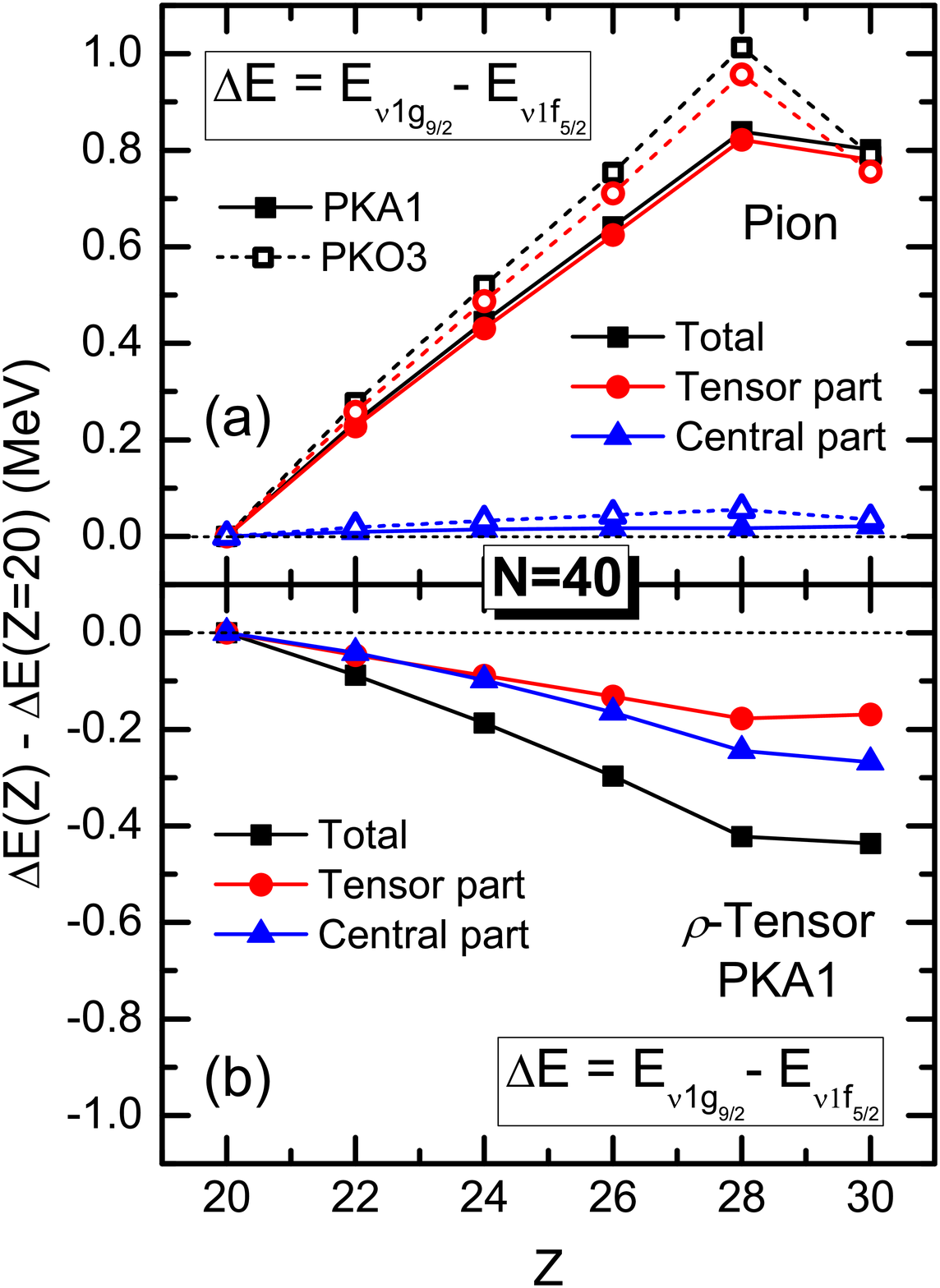}
\fi
\caption{ (Color online) Detailed contributions from the tensor and central parts of pion [plot (a)] and $\rho$-tensor [plot (b)] couplings to the energy difference $\Delta E = E_{\nu1g_{9/2}} - E_{\nu1f_{5/2}}$ between the neutron canonical single-particle states $\nu 1g_{9/2}$ and $\nu 1f_{5/2}$ along the isotonic chain of $N=40$. The results are extracted from the RHFB calculations with PKA1 (filled symbols) and PKO3 (open symbols).}\label{fig:rhopi}
\end{figure}

Similar systematical agreements and discrepancies can be also found in the relativistic cases. As seen from Fig. \ref{fig:fig1}(b) the calculations of RHB with DD-ME2 and RHFB with PKO2 present opposite isospin evolution behavior to the experimental trend, in which the isovector tensor component $\pi$-coupling is not taken into account. In fact due to the lack of isovector tensor couplings similar inconsistences are also found in other RHB calculations, e.g., with PKDD \cite{Long2004PRC}. Compared to those excluding isovector tensor components, the RHFB calculations with PKA1 present consistent trend as the experiments of the $\beta$-decay, which is mainly due to the tensor effects embedded in isovector $\pi$ and tensor $\rho$ couplings, particularly the former one. As shown in Fig. \ref{fig:fig2}, only the contributions from $\pi$ and tensor $\rho$-couplings are consistent with the experimental trend while the other channels present fairly remarkable cancellation with the referred behaviors. Consequently the calculations with PKA1 lead to fairly weak isospin dependence on the gap evolution of $N=40$ as compared to the non-relativistic results (SLy5+T and T31+T). In fact the $\pi$ contributions extracted from the calculations with PKO3 also present consistent trend with the experimental data [see Fig. \ref{fig:rhopi} (a)] while too strong cancellations are contributed by the other channels, leading to opposite isospin evolution in total [see Fig. \ref{fig:fig1} (b)].

It is interesting that the non-relativistic SHFB calculations with the tensor force and the relativistic RHFB ones including the $\pi$- and $\rho$-tensor couplings favor the same trend, and the conclusions are consistent with the experiments of $\beta$-decay \cite{Mueller1999PRL, Sorlin2001NPA} and semi-realistic mean fields calculations \cite{Nakada2010PRCR}. Different models may validate each other, which to some extent may indicate the reliability of the theoretical models and results. In addition, recent experimental \cite{Ljungvall2010PRC, Gade2010PRC} and theoretical \cite{Caurier2002EPJA, Lenzi2010PRC, Sato2012PRC} studies seem to indicate that a quantum phase transition from the spherical to deformed shapes takes place near $N=40$ for Cr and Fe isotopes, referred to as a {\it new island of inversion}. Similar to the one around $N=20$ \cite{Utsuno2002NPA, Giai2004PRC, Ren2006JPG}, the island of inversion of $N=40$ could be essentially concerned with the weakening semi-magicity of $N=40$ close to the neutron drip line, in which the tensor force plays a crucial role as seen from Figs. \ref{fig:fig1} and \ref{fig:fig2}.

To further clarify the tensor effects within RHFB theory, in Fig. \ref{fig:rhopi} are shown the contributions from the tensor and central parts of pion [plot (a)] and $\rho$-tensor [plot (b)] couplings to the evolutions of the energy difference $\Delta E_{gf} = E_{\nu1g_{9/2}} - E_{\nu1f_{5/2}}$, which corresponds to the gap of $N=40$ in PKA1 calculations. Here the tensor part of $\rho$-tensor coupling is extracted from the similar non-relativistic reduction as the $\pi$ pseudo-vector coupling \cite{Long2008EPL, Bouyssy1987PRC}. For the isovector $\pi$ coupling which in total presents consistent results with the experimental trend, its tensor part dominates such isospin dependence whereas the central part presents tiny effects in both PKA1 and PKO3 calculations. On the contrary, both the tensor and central parts of $\rho$-tensor coupling provide almost equivalent contributions to the isospin evolution of $\Delta E_{gf}$, opposite with the tensor $\pi$. It is also due to such distinct cancellation between the $\pi$ and $\rho$-tensor couplings that PKA1 provides consistent while fairly weak isospin dependence with the non-relativistic calculations of SLy5+T and T31+T.

In summary, the tensor effects in modeling the gap evolution of $N=40$ along the isotonic chain of $N=40$ are studied within the non-relativistic Skyrme-Hartree-Fock-Bogoliubov (SHFB) theory and the relativistic Hartree-Fock-Bogoliubov (RHFB) theory. The calculations with and without the isovector tensor component are compared with each other and the experimental trend. It is found that, when the tensor force is included, both the present non-relativistic and relativistic calculations yield similar trend for this gap evaluation and agrees with the experimental trend indicated by the $\beta$-decay studies, which to some extent may indicate the reliability of the theoretical results. Qualitatively the investigations indicate the necessity of the tensor force to interpret the experimental trend. In the RHFB calculations, both $\pi$-tensor and $\rho$-tensor couplings present significant contributions in the $T=0$ nucleon-nucleon interaction, which eventually determine the gap evolution of $N=40$, in agreement with the experimental trend. As some of the selected neutron-rich isotones may be deformed, the extension of the spherical mean-field calculations to the deformed one is planned as a future work to present a more comprehensive study.

This work was supported by the National Natural Science Foundation of China under Grant No. 11075066, the Fundamental Research Funds for Central Universities under Contracts No. lzujbky-2010-25 and No. lzujbky-2012-k07, and the Program for New Century Excellent Talents in University.



\end{document}